# Solid and hollow plasmonic nanoresonators for carrier envelope phase read-out

András Szenes[1,3], Dávid Vass[1,3], Balázs Bánhelyi[2,3], Péter Földi[4,5] Mária Csete[1,3,*]

[1]Department of Optics and Quantum Electronics, University of Szeged, 6720, Dóm tér 9, Szeged, Hungary
[2]Department of Computational Optimization, University of Szeged, 6720, Árpád tér 2, Szeged, Hungary
[3]HUN-REN Wigner Research Center of Physics, Konkoly-Thege Miklós út 29-33, Budapest, Hungary
[4]Department of Theoretical Physics, University of Szeged, Tisza Lajos krt. 84-86, H-6720 Szeged, Hungary
[5]ELI ALPS, ELI-HU Non-Profit Ltd., Wolfgang Sandner u. 3., H-6728, Szeged, Hungary

**Abstract**
The geometry of various plasmonic nanoantennae was numerically optimized to maximize their sensitivity to the carrier envelope phase (*CEP*) of the exciting ultra-short laser pulses. To verify the *CEP* sensitivity, the near-field response of the investigated nanoantennae was analyzed by combining frequency and time-domain numerical computations. The simulation methodology, capable of accurately calculating the time-dependent photocurrent stem from optical field emission, enabled the determination and optimization of all key parameters characterizing and governing the *CEP* dependence of the near-field responses. Three-types of structures were inspected, including individual triangular and teardrop-shaped nanoantennae and plasmonic lenses consisting of three hemispheres with gradually decreasing diameters. All structure types were optimized in solid and hollow compositions of gold nanoresonators as well. It was shown that hollow /solid singlets produce the largest /intermediate *CEP* dependent – to – *CEP* independent integrated current components' ratio, while their absolute *CEP* dependent integrated currents were the smallest /intermediate among the optimized structures. The highest /intermediate *CEP* sensitivity was achieved via solid plasmonic lenses due to their very large absolute *CEP* dependent integrated photocurrent originating from huge near-field enhancement in the nanogaps, while the integrated current components' ratio was smaller than for counterpart singlets.

**Introduction**
Metal nanoparticles support light-induced electron plasma oscillations on their surfaces, which phenomenon is known as localized surface plasmon resonance (LSPR) [1]. These localized plasmons confine the electromagnetic fields (**E**-field) to sub-wavelength volumes. This strongly below-diffraction-limited-confinement capability makes the plasmonic noble metal nanoparticles ideal for applications requiring simultaneously small mode volumes and enhanced **E**-field [2-14]. LSPRs are accompanied with inherent ohmic and radiative losses, leading to relatively small quality factors (*Q-factor*) and lifetimes, typically ranging from 10 fs to 100 fs [1].

The material, size and shape of the metal nanoparticles determine the resonance frequency, govern the reachable near-field enhancement (*NFE*), and the lifetime of plasmonic oscillations [15]. Through judicious optimization of the geometry, it is possible to maximize the *NFE* and to control the *Q-factor* for specific applications [16]. For example, supposing a given core-shell nanoparticle size, dipolar plasmonic resonances can be excited for two different metal shell thicknesses [17]. The LSPR on the thick-shell configuration possesses a lower *Q-factor*, accordingly it decays faster. Moreover, the plasmonic pulsed response can be even shorter, than the illuminating photonic pulse, which is well suited for applications, where few-cycle enhanced local **E**-field is necessary [14].

In the strong light-material interactions regime occurring on fs-to-attosecond time-scale, the photoelectrons emanating from the nanostructures are sensitive to the exact temporal evolution of the **E**-field, instead of being defined simply by the time-averaged **E**-field [18].

At high frequencies and sharp edges, spatially asymmetric **E**-field enhancement, hence photoelectron rectification is realizable that is controlled by the sub-cycle dynamics and mediately by the carrier envelope phase (*CEP*) of the exciting pulse [19-21]. Such rectified currents can be generated without complex structures and *CEP* detection can be performed on-chip in ambient conditions without a difficult lab-equipment [22].

Predesigned *CEP*-sensitive antennae can be used in experiments requiring *CEP* diagnostics [23], time-domain metrologies [24, 25], information processing and strong-field science [26, 27]. The bottleneck of using plasmonic nanoparticles is their relatively low damage threshold, which limits the applicable **E**-field strength and hence, the signal-to noise ratio as well as the ultimate *CEP* sensitivity [26, 27].

The resonant nanoparticles reduce the energy needed to generate few-cycle **E**-field due to the large *NFE* at their surface. Off-resonant particles enable to maintain the few-cycle nature of the **E**-field and *CEP* sensitivity, at the expense of compromised *NFE* [28]. Nanotriangles are commonly used due to their sharp apexes, which boost the *NFE* and generate well-defined hot spots acting as nano-scaled electron emission sources, while their asymmetric geometry is essential to register *CEP* sensitivity in the integrated current [27, 29]. With arrays of individual triangles that were tuned to ensure spectrally overlapping resonance with the near-IR pulse (1.177 μm), photoemission currents with 30 nA fundamental component and 1.5 pA *CEP*-sensitive first harmonic component were generated [27]. An order of magnitude current improvement was achieved using arrays constructed with off-resonant dimers of triangles compared to an array of singlet triangles [30]. The *CEP* sensitivity of properly tuned dimer arrays was further enhanced by using IR exciting pulses, namely 10-fold *NFE* and 3.3 electron per shot was reached at 2.7 μm wavelength [31].

The Fowler-Nordheim equation suitably describes the level of electron extraction from a metal, by including the effect of the barrier field [32, 33]. The theory is in good agreement with the experiments in case of optical field emission occurring at strong electric fields, corresponding to $\gamma\ll1$ Keldysh parameter [34, 35, 36, 37]. The Fowler-Nordheim equation constitutes an exponential term similar to that governing the atomic electron emission, and with a saddle point analysis this formula can be used to determine the photocurrents in a wide $\gamma$ interval as well [38, 39].

In the **E**-field strength dependence of the integrated photocurrent vanishing points were predicted and detected, where the *CEP* sensitivity suddenly drops to a significantly smaller, yet configuration dependent value [29]. The *CEP*-dependent photocurrent, numerically calculated using the Fowler-Nordheim equation as a function of incoming **E**-field amplitude, geometry and resonance wavelength, shows that a 50-100-fold *CEP* sensitivity improvement can be reached in the IR with respect to former results by properly predesigned triangular and teardrop-shaped antenna configurations [40].

The present study is focused on numerical optimization and investigation of the carrier-envelope phase sensitivity of gold nanoresonators illuminated by 795 nm central wavelength and 6.5 fs short-pulse, by balancing the trade-off between moderately prolonged near-field oscillation and improved **E**-field enhancement of resonant plasmonic nanoantennae, namely by considering both the *NFE* and *Q-factor* in the composite objective function.

**Methods**

To determine the photocurrent response of nanoantennae that are the subjects of short-pulse excitation, a comprehensive numerical methodology combining frequency and time-domain computations was developed. The simulations were conducted using the finite element method (FEM) based RF module of COMSOL Multiphysics, that ensures the accurate, full-wave analysis of the nanoantennae's response. The geometry of the nanoantennae was optimized with an in-house developed algorithm called GLOBAL [41, 42] to maximize the *NFE/Q-factor* ratio, considering the proportionality of the decay-time with the *Q-factor*. This objective function ensured large **E**-field enhancement, while maintaining few-cycled characteristics that is essential for high *CEP* sensitivity. Additional criterion was that the resonance frequency of the nanoantennae should be in the *FWHM* of the exciting pulse.

In the steady-state computations, continuous wave (CW) irradiation was applied to determine the *NFE* and absorption cross-section (*ACS*) spectra and to conclude about the maximal *NFE*.

In the time-domain computations, the optimized nanoparticle structures were excited by a cos$^2$-shaped pulse with a full-width-at-half-maximum (*FWHM*) of 6.5 fs and a central wavelength of 795 nm. The time-dependent photocurrent was calculated using the Fowler-Nordheim equation, considering the time-evolution of the local **E**-field (Fig. 1a, see Supplementary material for details).

The transient calculations were repeated for various *CEP* values, which enabled the characterization of the photocurrent and its temporal integral as a function of *CEP* and revealed the phase sensitivity of the nanoantenna near-field response. The photocurrent density computed using the Fowler-Nordheim equation ($J_{ph}(r,t)$) was integrated spatially on the nanonantennae's surface to determine the photocurrent ($I(t)$), and thereafter the photocurrent was integrated temporally through the duration of the exciting short-pulse to determine the corresponding amount of charges ($Q$). The *CEP*-independent fundamental component ($Q_0$, identifiable as a baseline, green dashed line in the inset of Fig. 1a) and *CEP*-dependent harmonic component ($Q_1$ manifesting itself in a sinusoidal modulation, blue solid line in the inset of Fig. 1a) of the integrated photocurrent were extracted in a wide **E**-field interval of [0.4, 2] V/nm. By analyzing the baseline and amplitude of the fitted sine curve, the most important quantities describing the impact of *CEP*, such as the $Q_1/Q_0$ ratio and the $Q_1^2/Q_0$ *CEP*-sensitivity were calculated [40]. These quantities provide the ratio of the *CEP*-dependent harmonic component of the integrated photocurrent with respect to the fundamental *CEP*-independent component and ultimately qualify the *CEP* sensitivity. The nanoresonators under investigation were triangular (TR) and teardrop (TD)-shaped nanoantennae and plasmonic lenses (LN), all made of gold and positioned onto an 8 nm ITO layer covered semi-infinite sapphire substrate. To exploit the potential advantage of core-shell compositions that can manifest itself in plasmonic field shortening, hollow (h) nanoresonators were also optimized aside their solid counterparts (s) (Fig. 1a). During the optimization procedure, the height of 20 nm was fixed, while the width and length of the nanoantennae as well as the shell thickness were varied independently, in order to maximize the *NFE/Q-factor* ratio defined as an objective function. In case of the hollow nanoparticles the minimal shell thickness was 5 nm.

**Results and Discussion**

Two relevant properties of the optimized nanoantennae are the spectrally resolved absorption cross-section (*ACS*) (Fig. 1b) and the maximal near-field enhancement (*NFE*) taken at the nanopantennae's apex and plasmonic lenses' gaps (Fig. 1c). These quantities provide information on the degree of detuning and strength of the plasmonic resonances (Table S1 and S2). The quality factor of the resonance can be determined also from the *FWHM* of the *ACS* spectra, which has a strong impact on the time-evolution and *CEP* sensitivity of the structures' response (Table S1) [16].

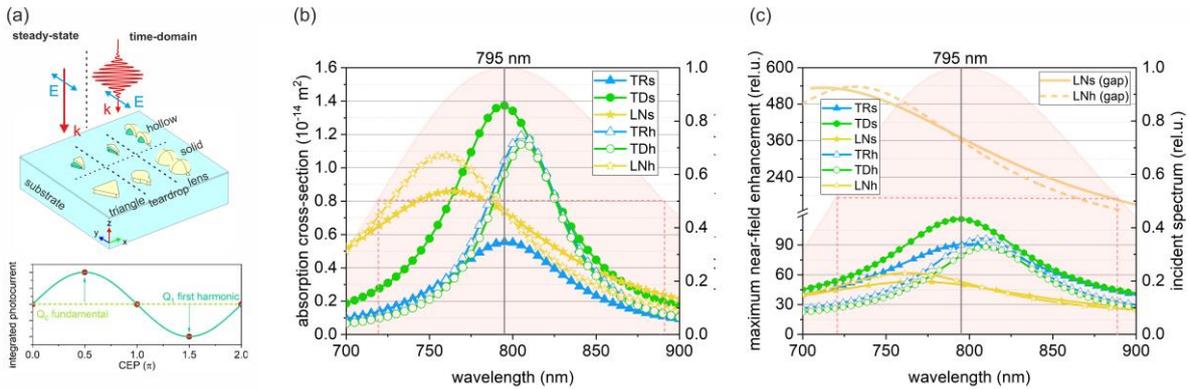

**Figure 1.** (a) Schematics of the optimized solid and hollow nanoantenna geometries and *CEP* dependent integrated photocurrents indicating the fundamental ($Q_0$) and first ($Q_1$) harmonics. Spectrum of the (b) absorption cross-section and (c) maximal near-field enhancement at the apex of nanoantennae and in the gap of plasmonic lenses. Shaded regions represent the incoming 6.5 fs pulse's spectrum, while the red dashed line indicates the *FWHM* of the spectrum. In the legend: TR – triangle, TD – teardrop, LN – lens, while the suffixes indicate s – solid, h – hollow.

For all optimized structures a single Lorentzian peak appears in the absorption cross-section (Fig. 1b) and maximal *NFE* (Fig. 1c) spectra within the *FWHM* of the pulse (Fig. 1b and 1c red dashed lines). This proves that every nanoantenna exhibits a localized plasmonic resonance around the central wavelength of the exciting pulse. The resonance is well-tuned to the central wavelength in the case of solid singlet nanoantennae. In contrast, the resonance peaks of the hollow singlets exhibit a small redshift, while in the case of plasmonic lenses a more pronounced blue-shift is observable (peak pos. in Table S1). These spectral shifts predict distinct spectral regions of optimal operation for the different inspected nanoantenna types (Fig. 1a and 1b).

The size of the nanoantennae was qualified by the geometric cross-section (*GCS*). The *GCS* of the optimized hollow nanoparticles is consistently smaller than the *GCS* of their solid counterparts (*GCS* in Table S1 and insets in Fig. 2). This size reduction compensates the redshift of the plasmon resonance that is introduced by the hollow geometry. According to the optimization via *NFE/Q*, it is advised to not-completely compensate the redshift through size reduction (Fig. 1b and 1c). The smaller size also results in a narrower spectrum. The geometric cross-section of the triangles is more than two-times larger than that of the teardrops, while the net *GCS* of plasmonic lenses is several times larger.

The hollow geometry is preferred in case of triangular nanoantenna, while larger *ACS* is achieved by solid composition in case of teardrops. Accordingly, the *ACS* of solid / hollow triangles is smaller / larger than the *ACS* of teardrop with analogous compositions. The solid and hollow plasmonic lenses exhibit similar *ACS* at the central wavelength of the incoming pulse. The achieved *ACS* of the plasmonic lens is intermediate / the smallest in case of solid / hollow composition compared to the singlet counterparts. The singlets exhibit larger *ACS* than *GCS*, while in case of plasmonic lenses the *ACS* is slightly smaller compared to the *GCS* (Table S1).

The *FWHMs* of the hollow nanoantennae is smaller than that of their solid counterpart. The triangles outperform the teardrops in the larger *FWHM* of *ACS* and *NFE* as well, both in solid and hollow nanoantennae. The plasmonic lenses exhibit the largest *FWHM* of *ACS* and *NFE* among the inspected optimized solid and hollow nanoantennae, providing the advantage of few-cycle properties preservation (*FWHM* values in Table S1 and Table S2). The broader spectral lines are due to the larger *GCS* of plasmonic lenses compared to the *GCS* of singlet nanoantennae.

The spectra of the maximal near-field enhancement at the apexes resemble the absorption cross-section spectra, however there are significant differences in the relative amplitudes of the two quantities. Higher *NFE* can be reached at the central wavelength of the pulse with the singlets of solid composition. The highest *NFE* is achieved with the solid teardrop-shaped antenna, which has the sharpest apex (upper insets in Fig. 2a and 2b, values taken at 795 in Table S2). In comparison, the hollow singlets produce lower *NFE,* because of the slight redshift of their resonance. Among the hollow nanoresonators, the highest *NFE* is achieved with the hollow triangular antenna (lower insets in Fig. 2a and 2b). The plasmonic lenses show significantly smaller *NFE* at the apex compared to the singlets. Slightly higher *NFE* can be reached at the central wavelength with the hollow composition, in contrast to singlets. The achieved small *NFE* can be partially attributed to the pronounced blue-shift of the resonance, but it is also limited by the confinement of near-field enhancement around the small nanogaps, rather than at terminating component acting as an apex (insets in Fig. 2c). In the nanogaps a large ~360-fold *NFE* is predicted with the steady-state calculations performed with large spatial resolution (yellow solid and dashed horizontal lines in Fig. 1c).

The maximum near-field enhancements are high enough to reach $\gamma<1$ Keldysh parameter, hence allow for entering into the optical field emission region for all inspected structures (Fig. S1a). A surface charge distribution corresponding to the time-instant representing the predominant dipolar distribution is shown in the insets of Figure 2. Based on these calculations the optimal singlet geometry is capable of sustaining dipolar plasmonic resonances.

In case of plasmonic lenses, the individual dipolar resonances are hybridized into a bonding antenna resonance. These charge distributions show, that the dipolar resonance is the most efficient to enhance *CEP* sensitive current.

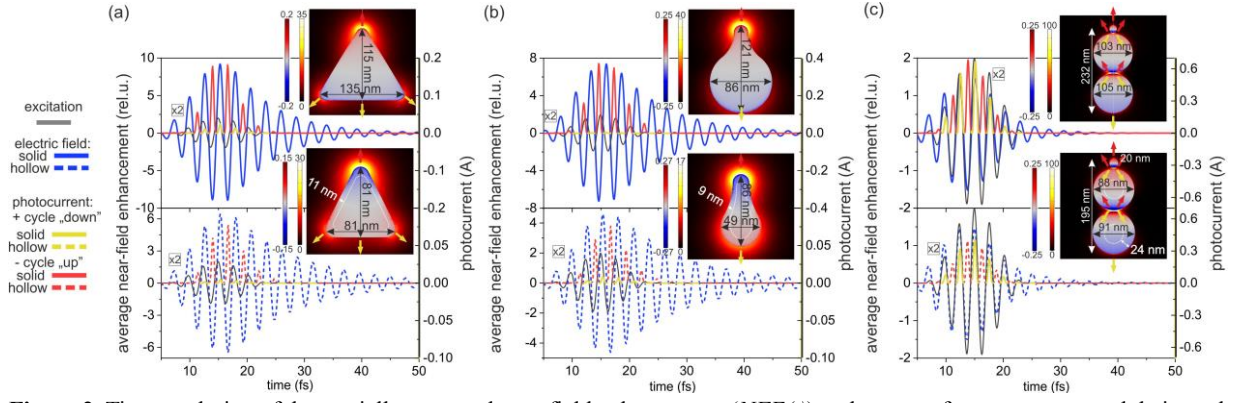

**Figure 2.** Time-evolution of the spatially averaged near-field enhancement (*NFE(t)*) at the apex of nanoantennae and their total photocurrent (*I(t)*) in the case of solid and hollow (a) triangle, (b) teardrop and (c) plasmonic lens.

The time-evolution of near-field enhancement (*NFE(t)*) averaged on the surface of apex is prolonged in case of singlets compared to the incoming short pulse, and the maximum in *NFE(t)* is significantly larger than unity indicating typical resonant nanoantennae (Fig. 2). The time-evolution of the plasmonic lenses is exceptional, as the solid plasmonic lens shortens, while the hollow plasmonic lens preserves the plasmonic response (decay in Table S3). Photocurrent spikes emerge around the time-instant corresponding to the peak values of the local **E**-field at the apexes. In Figure 2 yellow arrows indicate the photocurrent from the flatter side of nanoantennae (and from the bottom side of the half-spheres) during the positive half cycles of the excitation, while the red arrows indicate the photocurrent from the apexes (and from the top side of half-spheres) during the negative half cycles. Substantial differences arise in the amplitude and number of photocurrent peaks taken in the succeeding half duty-cycles, as a consequence of the asymmetry of each inspected nanoantenna's geometry ($N_{apex}$ and $N_{flat}$ in Table S3). The electron emission is maximized at the time instant, when the local **E**-field is anti-parallel with respect to the surface normal.

The decay of the enhanced near-field around the solid and hollow triangles is significantly prolonged compared to the incident exciting 6.5 fs pulse (Fig. 2a). The lifetimes of the plasmons are 10.2 fs and 16.9 fs for solid and hollow configuration respectively, indicating a plasmon resonance of larger *Q-factor* for the latter, in accordance with the steady-state computations. Although, their maximal time-averaged steady-state *NFE* values are close to each other, namely it is 90.9 / 95.5-fold in case of the solid / hollow triangles (Fig. 1c, steady-state), the maximum in *NFE(t)* at the apex is 1.4-times larger for solid configurations. The corresponding peak in the *NFE(t)* is 9.2-fold for the solid and 6.4-fold for the hollow configuration, which is suitable for large photocurrent generation (*NFE* values in Table S3). Around the global maximum in *NFE(t)*, large photocurrent peaks appear during the negative **E**-field half-cycles, i.e. from the sharp apex of the nanotriangles. The smaller number of detectable peaks are significantly smaller in the positive half-cycles, as negligible photocurrent emanates from the flat, bottom side of nanotriangles due to the asymmetric antenna geometry and local **E**-field around the triangles ($N_{apex}$ and $N_{flat}$ in Table S3). The global maximum of the photocurrent is significantly larger (more than doubled) for the solid configuration ($I_{max}$ in Table S3,), moreover the photocurrent from the flat side is significantly smaller for the hollow configuration. This predicts that the *CEP* sensitive current could be more efficiently detectable in solid geometry, however both compositions are advantageous to attain large *CEP* sensitivity (Fig. 2a).

Relatively long resonances are sustained on the teardrop geometry as well (Fig. 2b). The lifetimes of 11.4 fs of solid teardrop and 12.9 fs of hollow teardrop are larger and smaller with respect to lifetimes observed in case of the solid and hollow triangles, respectively (Fig. 1c, Table S3). The maximal steady-state *NFE* is larger / smaller for the solid / hollow teardrop compared to the triangular counterparts, i.e. the solid teardrop's advantage is more pronounced (Fig. 1c, Table S2). Larger maximum is attainable in *NFE(t)* with the solid configuration (7.4-fold) than with the hollow one (4.6-fold), (*NFE* values in Table S3), however both teardrop compositions possess smaller maximum in the *NFE(t),* than the triangular nanoantennae of analogous composition.

In case of solid teardrop the generated reached $I_{max}$ photocurrent is more than 2-times larger compared to the solid triangle, according to the large maximal steady-state *NFE* and despite the smaller maximum in *NFE(t)* ($I_{max}$ in Table S3). In contrast, the hollow teardrop shows almost the same photocurrent, as the hollow triangle, although both the maximal time-averaged *NFE* and the maximum in *NFE(t)* are smaller. Accordingly, the $I_{max}$ is significantly larger on solid TD than on its hollow counterpart. Despite the relatively large photocurrents during the negative half-cycles, the smaller number of detectable photocurrent peaks are smaller during the positive half-cycles on the teardrops, which is beneficial in the *CEP*-sensitivity, especially for the solid composition (Fig. 2b, $N_{apex}$ and $N_{flat}$ in Table S3).

The maximal steady-state *NFE* at the apex of the plasmonic lenses is significantly smaller than in case of the singlet structures (Fig. 1c, Table S2). Moreover, the lifetime of the plasmon is only 5.9 fs and 6.5 fs in case of solid and hollow plasmonic lenses, respectively, in accordance with their small *Q-factor*. Based on these lifetimes, the plasmonic lenses are suitable in preserving, or even shortening the pulse-length of the few-cycled **E**-field compared to the incoming short pulse (Fig. 1c). Although, the maximum in *NFE(t)* is relatively small (1.5-fold for both compositions) (*NFE* values in Table S3), the maximum in the photocurrent is significantly larger than in case of either singlets ($I_{max}$ in Table S3 and insets in Fig. 2c). This is achieved, though there is significant photocurrent for both halves of each duty cycles, according to the more symmetric nature of the nanogaps and due to the huge *NFE* at the gaps compared to the environment of singlets, ($N_{apex}$ vs $N_{flat}$ in Table S3). The $I_{max}$ is larger on the solid plasmonic lens than on its hollow counterpart, but the maximum values are more comparable than in case of singlets.

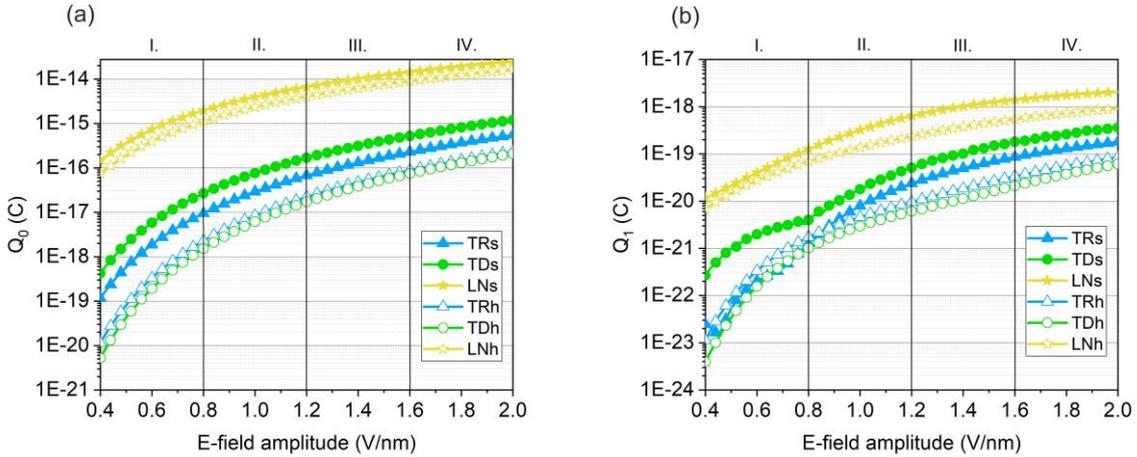

**Figure 3.** Integrated photocurrent characteristics of optimized configurations. (a) Fundamental ($Q_0$) and (b) first harmonic ($Q_1$) of the integrated photocurrent as a function of incident electric field amplitude. In the legend: TR – triangle, TD – teardrop, LN – lens, while the suffixes indicate s – solid, h – hollow.

Considering the modulations of the integrated photocurrents, their *CEP* dependent component and their respective ratios, the inspected **E**-field interval was separated into four different regions, to analyze them thoroughly. Accordingly, the preferred structure types were compared in Intervals of I / II / III / IV: 0.4-0.8 V×nm$^{-1}$ / 0.8-1.2 V×nm$^{-1}$ / 1.2-1.6 V×nm$^{-1}$ / 1.6-2.0 V×nm$^{-1}$ (Fig. 3 and Fig. 4).

The integrated *CEP*-independent integrated photocurrent component ($Q_0$) nonlinearly increases with the exciting **E**-field amplitude, however the rate of the increase gradually decreases (Fig. 3a). Both the nonlinear tendency and the decreasing slope are the direct consequence of the Fowler-Nordheim equation (Fig. S1b). Due to the nonlinear growth, it is possible to extract an integrated photocurrent in an interval spanning six orders of magnitude - ranging from 10$^{-20}$ to 10$^{-14}$ C - with the six examined structure types (Fig. 3a and Table S4). This is the result of a calculation on currents of a single nanoantenna and plasmonic lens, hence the net current could be significantly scaled up by using an array of them and increasing the target size. The largest slope can be achieved with the solid lens, followed closely with its hollow counterpart in the I-interval ($Q_0$ in Table S4). Accordingly, the largest differences between $Q_0$ values are observable at the smallest **E**-field strengths. By using different configurations of nanoantennae up to 4 orders of magnitude $Q_0$ modulation span is achievable in the I-interval.

The solid compositions show larger integrated photocurrents for all nanoantenna types in all inspected **E**-field intervals ($Q_0$ in Table S4). Approximately three-times and one order of magnitude larger $Q_0$ is achieved in optimized solid triangular and teardrop structures compared to their hollow counterparts respectively, according to their larger *NFE* (Table S2 and S3). Almost two-times larger integrated photocurrent can be achieved with solid plasmonic lens compared to hollow one, despite the comparable steady-state maximal *NFE* and equal maxima in *NFE(t)* (Table S2 and S3). In the case of plasmonic lenses considering only the maximal *NFE,* calculated at the terminating components acting as apexes, one could underestimate the integrated photocurrent. However, huge *NFE* confinement and large integrated photocurrent are achieved in the nanogaps.

The $Q_0$ of singlets is at least one / two orders of magnitude smaller but can be even three / four orders of magnitude smaller, than in case of solid / hollow plasmonic lens in the inspected **E**-field interval.

The large achievable integrated photocurrents can be explained by the ~360-fold enhancement in the 2 nm nanogaps (Fig. 1c). If one intends to achieve large $Q_0$, plasmonic lenses are preferable due to the enormous *NFE* and the resulted more intense electron escape near the nanogaps. *The $Q_0$* correlates with the maximum in the photocurrent (Fig. 3a to Fig. 2, Table S6), however, it correlates with the maximal steady-state *NFE* at the apex only in case of singlet nanostructures (Table S2 and Table S4). The achieved $Q_0$ values show strong correlation with the maximal *NFE* including the nanogaps in all intervals and anticorrelates with the *Q-factors* (Table S6).

The *CEP* sensitive integrated photocurrents ($Q_1$) non-monotonously increase with the **E**-field amplitude (Fig. 3b). Shallow vanishing points (VP) appear in distinctly different **E**-field strength regions for different nanoantenna geometries, however dominantly in the first two inspected amplitude intervals. The $Q_1$ is 3-4 orders of magnitude smaller than the $Q_0$ ($Q_1/Q_0$ in Table S5) and modifies between $10^{-23}$ – $10^{-18}$ C, depending on the short-pulse intensity and **E**-field strength as well as the antenna configuration (Fig. 3b, $Q_1$ in Table S4). Based on the amplitude of the *CEP*-sensitive $Q_1$ both for individual nanoantennae and for plasmonic lenses, the solid geometry is preferable dominantly (except the triangles in the Interval I). The $Q_1$ of singlets is at least one order of magnitude smaller than in case of plasmonic lenses of the same composition in the inspected **E**-field intervals, three / four orders of magnitude smaller relations are achieved in case of solid and hollow TRI / TD at the beginning of interval I.

Apart from the VPs at small **E**-field strengths in case of the solid triangle and teardrop, the $Q_1$ completely correlates with the $Q_0$ and $I_{max}$ in the optimized structures (Table S6). It is an additional proof about that all nanonatenna types are equally optimized. The $Q_1$ correlates with the maximal *NFE* including the nanogaps and anticorrelates with the *Q-factors,* as it is expected from the strong correlation between the *CEP* insensitive ($Q_0$) and sensitive ($Q_1$) components of the integrated photocurrent (Table S6).

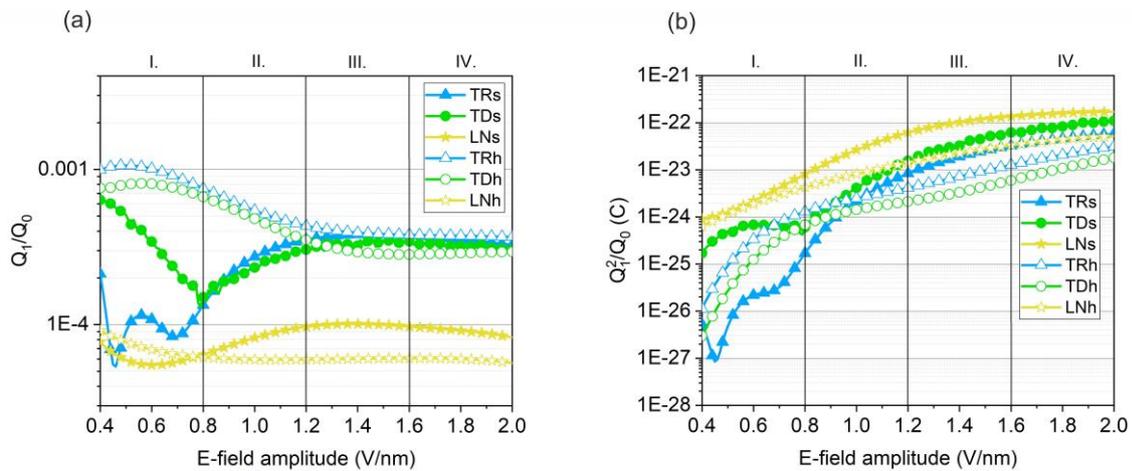

**Figure 4.** Carrier envelope phase sensitivity characteristics of optimized configurations. (a) $Q_1/Q_0$ ratio and (b) $Q_1^2/Q_0$ sensitivity as a function of incident electric field amplitude. In the legend: TR – triangle, TD – teardrop, LN – lens, while the suffixes indicate s – solid, h – hollow.

Although, $Q_0$ and $Q_1$ themselves carry useful information, the quantities derived from $Q_0$ and $Q_1$ provide the reasonable evaluation of the *CEP* sensitivity of the optimized structures. The $Q_1/Q_0$ shows the relative strength of *CEP* sensitive integrated photocurrent to the *CEP* independent signal (Fig. 4a). It is a non-monotonously varying quantity throughout the inspected **E**-field intervals and may increase or decrease in a specific **E**-field interval depending on the nanoantenna type. The strong modulations are the direct fingerprints of the VPs, while shallow modulations originate from the different slopes of the $Q_0$ and $Q_1$ amplitude - **E**-field tendencies.

The $Q_1/Q_0$ is modulated within one order of magnitude in all configurations in the complete inspected **E**-field amplitude interval ($Q_1/Q_0$ in Table S5). The tendencies strongly depend on the specific nanoantenna type. The singlets behave completely differently in Intervals I-II and III-IV. In the former two, the solid singlets have one or even multiple VPs, accordingly the hollow singlets outperform the solid ones in the first half of the inspected **E**-field region ($Q_1/Q_0$ Interval I-II in Table S5).

The relations are nearly interchanged in Intervals III and IV, where the hollow singlets have very shallow VPs, however the hollow triangle is still the most preferable structure considering the $Q_1/Q_0$ ratio (except the termination of the inspected interval, where the ratio is the same as for the solid TR) ($Q_1/Q_0$ in Interval III-IV in Table S5).

The plasmonic lenses possess the smallest $Q_1/Q_0$ ratio, i.e. the *CEP* dependent component of the total integrated photocurrent is relatively small compared to the *CEP* independent component ($Q_1/Q_0$ in Table S5). This can be explained by the composite geometry of the plasmonic lenses. Although, the complete geometry of the plasmonic lenses is asymmetric, most of the electron emission originates from the nanogaps, which produce large **E**-field enhancement in both halves of each duty-cycle (Fig. 2c). This inherent symmetry at the nanogaps reduces the *CEP* sensitivity of $Q_1/Q_0$ ratio. However, due to the large emanating current accompanied with a small *Q-factor*, the plasmonic lenses are still very competitive. The hollow plasmonic lens outperforms its solid counterpart only in Interval I, where a VP appears in case of the solid plasmonic lens ($Q_1/Q_0$ in Interval I in Table S5).

The $Q_1/Q_0$ shows strong anticorrelation with the maximal *NFE* including the nanogaps and the amplitude of photocurrent peaks and indicates a strong correlation with the *Q-factors* of the resonances at play, especially in Intervals I and II (Table S7). The $NFE_{max}$ and $I_{max}$ anticorrelations are weaker in regions corresponding to smaller **E**-field strengths, because of the modulations caused by the appearance of the VPs. The correlations are reversed compared to $Q_0$ and $Q_1$, as the $Q_1/Q_0$ ratio anticorrelates with the constituting quantities (Table S7).

The *CEP* sensitivity is usually determined by the ratio of $Q_1^2/Q_0$ (Fig. 4b). It is non-monotonously increasing throughout the inspected **E**-field interval. The modulations are the direct consequences of the existence of the VPs. The sensitivity varies between $10^{-28}$ and $2\times10^{-22}$ C ($Q_1^2/Q_0$ Table S5) and for each specific configuration the modulation covers 2-4 orders of magnitude.

The hollow teardrop and triangle show relatively small sensitivity because of their small $Q_1$, though the hollow triangle is preferable exceptionally in Interval I of the smallest **E**-field strengths, where no VPs appear for hollow nanoantennae. The singlets are always less preferable than the plasmonic lenses of the same composition, in all inspected intervals. Although, the solid teardrop becomes better than the hollow plasmonic lens in Interval II, III and IV, and the solid triangle also outperforms it in Interval IV (Fig. 4b, $Q_1^2/Q_0$ in Table S5).

Despite the small $Q_1/Q_0$, the solid plasmonic lens shows the largest sensitivity among the optimized structures (Fig. 4b, $Q_1^2/Q_0$ in Table S5,). It outperforms the sensitivity of the hollow plasmonic lenses in all **E**-field regions due to its large *CEP* dependent current density ($Q_1^2/Q_0$ in Table S5). Moreover, also the hollow plasmonic lens outperforms both types of the singlets in the Intervals I and II, except the solid teardrop in the Interval II.

The $Q_1^2/Q_0$ correlates with the constituting quantities, hence the sensitivity correlates with the maximal *NFE* including the nanogaps and the maximum in the integrated photocurrent, while anticorrelations can be recognized with the *Q-factors* and the $Q_1/Q_0$ ratio. Due to the existence of vanishing points the (anti)correlations are weaker in Intervals III and IV (Table S7). Better correlations are achieved with $NFE_{max}$ and $Q_0$ / $Q_1$ / $I_{max}$ in Interval I / I-III / II-IV.

## Conclusion

In conclusion, it is demonstrated that the plasmonic resonances on metal nanoantennae offer intriguing possibilities for achieving enhanced electromagnetic field confinement at the nanoscale and controlling the time-evolution of the few-cycle plasmonic field at fs-time-scale.

Plasmonic nanoantenna singlets, as solid and hollow triangles and teardrops, are preferable from the point of view of the achievable large $Q_1/Q_0$ ratio originating from their strongly asymmetric near-field enhancement. Among them, the hollow singlet nanoantennae are the most preferable in terms of this ratio, as they show the strongest *CEP* dependent integrated photocurrent component, especially in the smaller **E**-field amplitude region. Because of the small *CEP* sensitive integrated photocurrent component, their smaller $Q_1^2/Q_0$ sensitivity is comparable only in these smaller amplitude regions with their solid counterparts. Only the $Q_1^2/Q_0$ of the hollow TR outperforms the sensitivity of the solid TR in the smallest **E**-field interval.

The analysis of various optimized plasmonic nanoantennae has shown that solid plasmonic lenses ensure the largest *CEP* sensitivity. Their advantage originates from the large electron emission rate from the huge near-field enhancement in the nanogaps. Although, the $Q_1/Q_0$ *ratio* is relatively small compared to singlets, because of the symmetry of the nanogaps, the large *CEP* dependent $Q_1$ component results in an extraordinary sensitivity. Larger $Q_0/Q_1$ ratio (except in the Interval I) and $Q_1^2/Q_0$ sensitivity in all Intervals can be attained with a solid lens than with its hollow counterpart.

Vanishing points have an impact on the *CEP* dependent and independent integrated photocurrent components' $Q_1/Q_0$ ratio and $Q_1^2/Q_0$ sensitivity that can be controlled, and the smallest impact is attainable in case of plasmonic lenses, except the ratio at small **E**-field strength. By alternating the nanoantenna configuration, the VPs can be tuned properly, thus allowing *CEP* monitoring through wide **E**-field regions.

## Associated content

Detailed information on the near-field response of optimized nanoantennae, validation. CEP independent and dependent integrated photocurrent components arranged in tables (PDF format).

## Corresponding Author

*Mária Csete, Department of Optics and Quantum Electronics, University of Szeged, Dóm tér 9, Szeged 6720, Hungary, Email: mcsete@physx.u-szeged.hu

## Author Contributions

M.C. proposed the concept of individual plasmonic nanoantennae and plasmonic lens configurations' optimization to maximize CEP sensitivity. B.B. provided the method of numerical optimization. A.S. and V.D. contributed to data collection and analysis of the optimized nanoresonator configurations. A.S. wrote the first version of the manuscript and created the visualizations, M.C., V.D. and F.P. contributed to the theoretical explanation, writing and revision of the manuscript. All authors have discussed the results, commented on the manuscript and given approval to the final version of the manuscript.

## Funding Sources

This work was supported by the National Research, Development and Innovation Office (NKFIH) of Hungary, projects: "Optimized nanoplasmonics" (K116362), "Nanoplasmonic Laser Inertial Fusion Research Laboratory" (NKFIH-2022-2.1.1-NL-2022-00002) and "National Laboratory for Cooperative Technologies" (NKFIH-2022-2.1.1-NL-2022-00012) in the framework of the Hungarian National Laboratory program. Supported by the UNKP-20-3 new national excellence program of the ministry for innovation and technology from the source of the National Research, Development and Innovation Fund.

## Acknowledgements

Partial support by the ELI-ALPS project is acknowledged. The ELI-ALPS project (GINOP-2.3.6-15-2015- 00001) is supported by the European Union and cofinanced by the European Regional Development Fund.

## Notes

The authors declare no competing financial interest.